\documentclass [12pt]{article}
\usepackage{geometry} 
\usepackage[english]{babel}
\usepackage{graphicx}

\begin{document}

\textbf{Energy dependence of cross section of photonuclear
reactions on indium isotopes}

\begin{center}
V. I. Zhaba, I. I. Haysak, A. M. Parlag, V. S. Bohinyuk, M. M.
Lazorka
\end{center}

\begin{center}
Uzhgorod National University, 88000, Uzhgorod, Voloshin Str., 54
\end{center}

\textbf{Abstract}

Experimental isomeric yield ratios for the $^{113}$In($\gamma
$,n)$^{112m,g}$In reactions on the betatron B25/30 bremsstrahlung
gamma beam of energy range 12-25 MeV are measured. Effective
cross-sections of ($\gamma $,n)-reactions with $^{112m}$In and
$^{114m}$In isomers output are calculated. The Penfold-Leiss and
Tikhonov methods are applied to solve the Volterra integral
equation. The obtained experimental cross-sections are compared
with theoretical calculations using the TALYS-1.6 code.



\textbf{1. Introduction}

The purpose of the majority of physical researches that are
carried spent on brake beams electronic accelerators, - studying
of power dependence of cross sections of different photonuclear
processes. As the spectrum received from the accelerator $\gamma
$-quantum's has continuous character, therefore in experiment not
cross section is measure of reaction, it is so-called output. The
output is intensity photonuclear to the process, the doze
attributed to unit $\gamma $-quantum's that have passed through a
target with researched substance at various values of the top
border of a brake spectrum \cite{1,2,3}.

The yield of reaction is directly connected with effective cross
section of reaction by integral equation of the first kind:

\begin{equation}
\label{eq1} Y(E_{\gamma \max } ) = \int\limits_{E_m }^{E_{\gamma
\max } } {\sigma (E)\Phi (E,E_{\gamma \max } )dE} ,
\end{equation}

where $ E_{m}$ and \textit{$\sigma $(E)} are the threshold and the
cross section of a reaction, $E_{\gamma \max }$ is the maximal
energy of a bremsstrahlung $\gamma $-spectrum, \textit{ $\Phi
$(E,E}$_{\gamma \max })$ is an energy $E$ of spectrum of $\gamma
$-quanta (Shiff's spectrum \cite{4}).

The cross section of the reaction can be received from
experimental data about yield as a result of the solution of an
inverse problem (\ref{eq1}). For the numerical solution of this
integral equation different mathematical methods were developed.
The most widespread methods are ``a difference of photons'', ``the
least structure'' Cook's method \cite{5}, ``an inverse matrix''
(Penfold-Leiss method) \cite{6}, ``regularizations'' (Tikhonov
method) \cite{7}. Penfold-Leiss and Tikhonov methods have
different forms of the effective photon spectrum - the hardware
function of the method. Except the direct solution of the inverse
problem there are other methods for determination of information
about the cross section, namely, a combination of reaction outputs
and a method of reduction.

The conditions of well-posed problem given by Hadamar \cite{8}
should have the properties: 1) a solution exists in space of
possible values; 2) the solution should be the unique; 3) the
solution's behavior changes continuously with the initial
conditions.

This paper deals with the determination of the experimental cross
sections of ($\gamma $,n)- reaction on isotopes of indium with the
formation of isomers $^{112m}$In and $^{114m}$In.

\textbf{2. Cross section of reaction} $^{ 113}$\textbf{In($\gamma
$,n)}$^{112m}$\textbf{In}

With the bremsstrahlung gamma beam of the betatron B25/30 (UzhNU,
Uzhgorod) there was measured isomeric ratios of the yields for the
reaction $^{113}$In($\gamma $,n)$^{112mg}$In. The energy of
electron beam was changed in the interval 12-25~MeV with a step of
1~MeV. Isomeric state of the nucleus $^{112m}$In decays with a
half-life of 20.9~m, emitting $\gamma $-rays with energy 155~keV
(the quantum yield is 13{\%}). $\beta ^{ + }$- decay of the ground
state (T$_{1 / 2}$=14.4~m) is accompanied by the emission of
$\gamma $-rays with energies 511, 606 and 618~keV, with the
quantum yields 44, 1.2, 5.3{\%}, respectively. Time of irradiation
of targets and time for measurement were 10-20~m, and time for
cooling was 5-10~m.

Nuclear-physical parameters of the isomers $^{112m}$In and
$^{114m}$In are specified in Table 1, where $E_{m}$ is a threshold
($\gamma $,n)-reaction, $T_{1/2}$ is a half-life period;
$E_{\gamma}$ is energy of $\gamma $-ray that radiates the isomer;
$J_{m},$ and $J_{g}$ are the total angular momentum of the
isomeric and the ground state; "+" or "-" is the state parity.

Table 1. Nuclear-physical characteristics of investigated isomers

\begin{tabular}{|l|l|l|l|l|l|}
\hline Isomer& $E_{m}$, \par MeV& $T_{1/2}$& $E_{\gamma}$, keV&
$J_m^P $& $J_g^P $ \\
\hline $^{112m}$In& 9.58& 20.9 m& 155& 4$^{+}$& 1$^{ + }$ \\
\hline $^{114m}$In& 9.03& 43 ms& 310& 8$^{-}$& 1$^{ + }$ \\
\hline
\end{tabular}

Experimental results for yields ratio of reaction
$^{113}$In($\gamma $,n)$^{112mg}$In are shown in Table 2.

According to the obtained isomeric ratio of yields Y$_{m}$/Y$_{g}$
it is possible to calculate the cross sections of isomeric state
excitation in the reaction $^{113}$In($\gamma $,n)$^{112m,g}$In.
For this we need to use the known values of total cross section of
($\gamma $,n)- reaction \cite{9}. We used the connection between
the yields and the integral cross-sections $\sigma _{int}$:

\begin{equation}
\label{eq2} K(E_{\gamma \max } ) = \frac{Y^{{
}^{113m}In}(Е_{\gamma \max } )}{Y^{{ }^{113g}In}(Е_{\gamma \max }
)} \cong \frac{\sigma _{int}^{{ }^{113m}In} (Е_{\gamma \max }
)}{\sigma _{int}^{{ }^{113g}In} (Е_{\gamma \max } )},
\end{equation}

\begin{equation}
\label{eq3} \sigma _{int} = \int\limits_{E_p }^{E_{пор} } {\sigma
(E_\gamma )d} E_\gamma = \frac{Y(E_{\max } )(E_{\max } - E_m
)}{\int\limits_{E_p }^{E_{пор} } {\Phi (E,E_{\gamma \max } )d}
E_\gamma },
\end{equation}

where $E_{m}$ -- energy of isomeric level.

Table 2. Isomeric ratios $Y_{m}/Y_{g}$ of reaction
$^{113}$In($\gamma $,n)$^{112mg}$In

\begin{tabular}{|l|l|l|}
\hline E, MeV& Y$_{m}$/Y$_{g}$& $\Delta $(Y$_{m}$/Y$_{g})$ \\
\hline 12&1.86&0.32 \\
\hline 13&2.34&0.30 \\
\hline 14&2.52&0.21 \\
\hline 15&2.91&0.12 \\
\hline 16&2.83&0.17 \\
\hline 17&2.76&0.08 \\
\hline 18&2.62&0.10 \\
\hline 19&2.44&0.05 \\
\hline 20&2.6&0.04 \\
\hline 21&2.39&0.12 \\
\hline 22&2.48&0.05 \\
\hline 23&2.41&0.06 \\
\hline 24&2.45&0.08 \\
\hline 25&2.51&0.06 \\
\hline
\end{tabular}

It should be noted that for solving the integral equation
(\ref{eq1}) by Tikhonov's method ones use Shiff's spectrum
\textit{ $\Phi $(E,E}$_{\gamma \max })$ as a core of integral
equation. In Fig. 1 the cross sections obtained by Penfold-Leiss
(PL) and Tikhonov's (T) methods are shown. The difference between
them is 4-5{\%}.

\pdfximage width 177mm {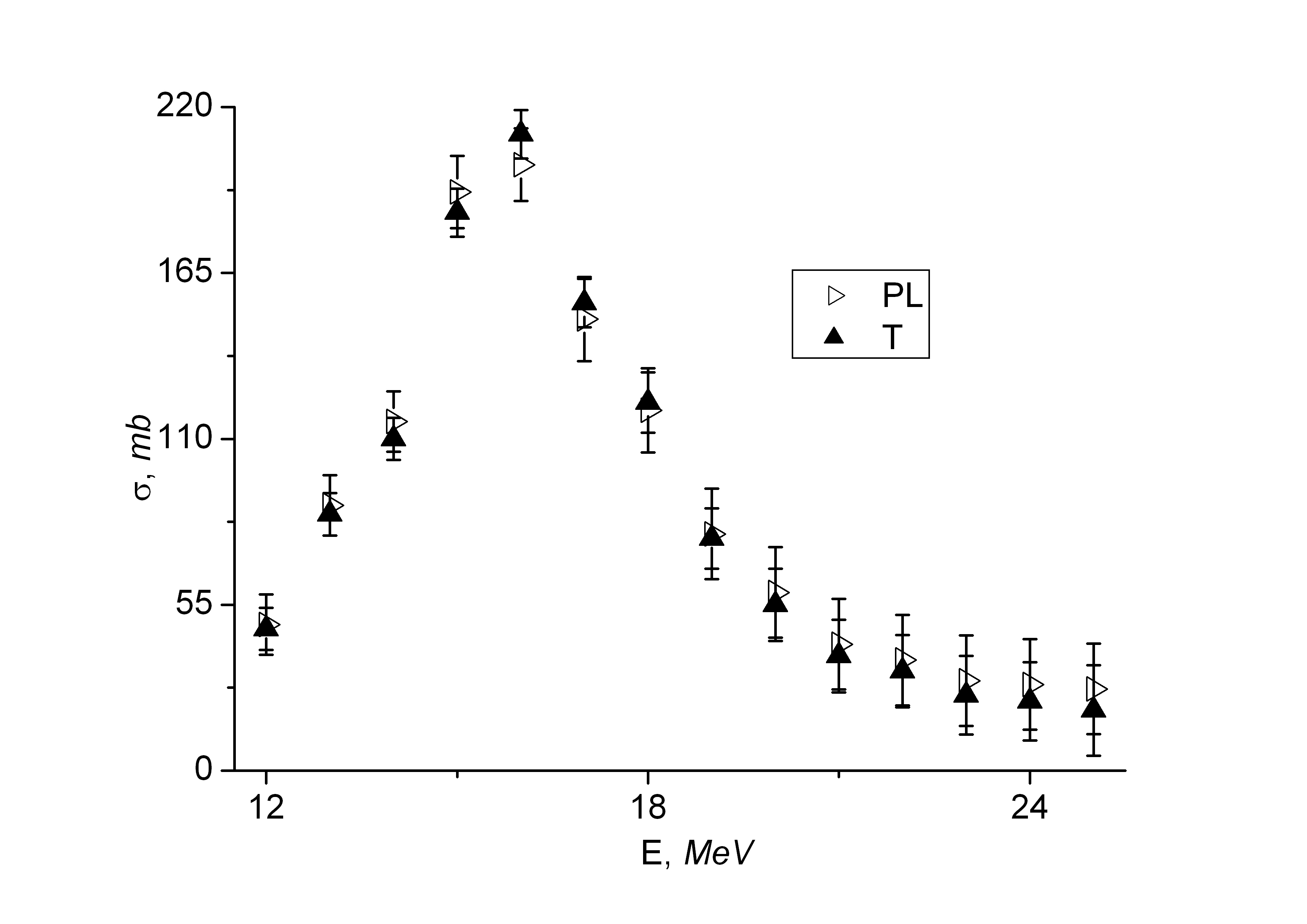}\pdfrefximage\pdflastximage

Fig.~1. Energy dependence of cross section of reaction
$^{113}$In($\gamma $,n)$^{112m}$In

It is possible to use package TALYS-1.6 \cite{10,11} for the
simulation of nuclear reaction $^{113}$In($\gamma $,n)$^{112m}$In.
When calculating cross sections in TALYS-1.6 the density level
model was selected for a given nuclide (ldmodel parameter). In the
package there are three phenomenological level-density models and
two variants for microscopic level densities. In particular:
ldmodel~1 is Fermi-gas model with a constant temperature;
ldmodel~2 is Fermi-gas model with reverse shift; ldmodel~3 is
generalized superfluidity model; ldmodel~4 is microscopic level
densities from Goriely's table \cite{10}; ldmodel~5 is microscopic
level densities from Hilaire's table \cite{10}.

On the consolidated Fig.~2. our experimental data of cross section
for reaction $^{113}$In($\gamma $,n)$^{112m}$In (Fig.~1) and
experimental data \cite{12} are shown in comparison with
theoretical calculations in the package TALYS-1.6.

\pdfximage width 177mm {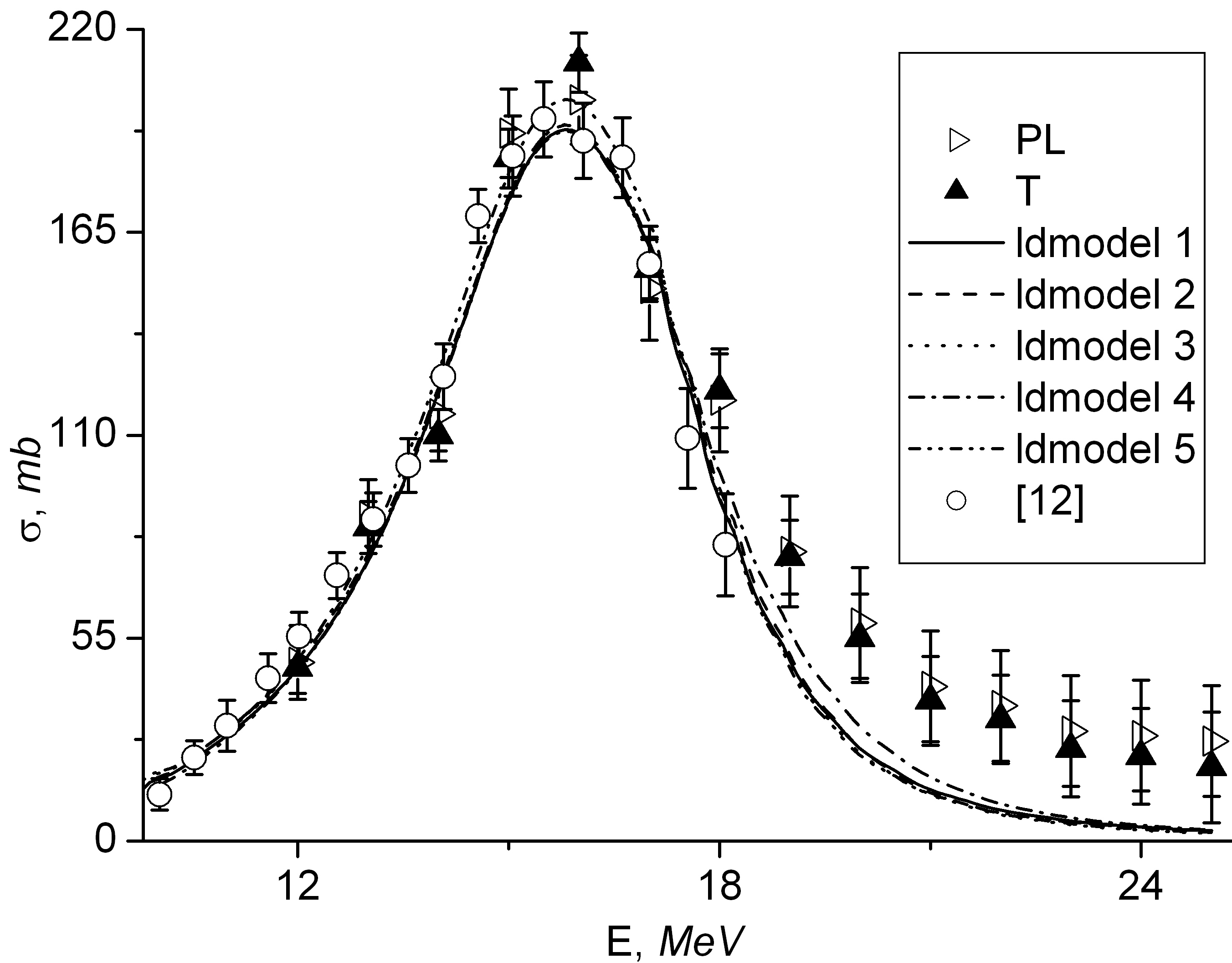}\pdfrefximage\pdflastximage

Fig.~2. Energy dependence of cross section of reaction
$^{113}$In($\gamma $,n)$^{112m}$In

The results of fitting of the calculated cross section peaks for
the reaction $^{113}$In($\gamma $,n)$^{112m}$In in the package
TALYS-1.6 are shown in Table 3, where the following designations
are used: \textit{$\chi $}$^{2}$ per degree of freedom of
function; $S$ is the area under the peak in the MeV*mb; $\Gamma $
is the full width at half maximum; ldm-1(\ref{eq5}) are the
numbers of the density level model for the nuclide. As an
approximating function the Gauss function was chosen

\begin{equation}
\label{eq4} y = y_0 + \frac{A}{w\sqrt {\pi / 2} }e^{{ - 2(x - x_c
)^2} \mathord{\left/ {\vphantom {{ - 2(x - x_c )^2} {w^2}}}
\right. \kern-\nulldelimiterspace} {w^2}}.
\end{equation}

Additionally it was fitting the same picks by the Breit-Wigner
function (Table 4):

\begin{equation}
\label{eq5} y = y_0 + \frac{2A}{\pi }\frac{w}{4(x - x_C )^2 +
w^2},
\end{equation}

obtained parameters of which are given in Table 5. Analyzing the
data of Tables 3 and 4, it is seen that adjustment of the peak by
the Breit-Wigner formula (\ref{eq5}) gives better results than
Gaussian by the formula (\ref{eq4}).

Consequently, the experimental results for the cross section of
the reaction $^{113}$In($\gamma $,n)$^{112m}$In matches with the
calculated one in the package TALYS-1.6 in the region of maximum.

For energies of 18-25~MeV there is some difference between the
results of theory and experiment. It can be explained by the fact
that the TALYS-1.6 is not included nappy processes, but reaction
$^{113}$In($\gamma $,2n)$^{112}$In contributes to experimental
results, which begins with the threshold 16.3~MeV. Some
calculation results of the above reactions is given in \cite{13}.

Table 3. The results of processing the peaks of cross section for
reaction $^{113}$In($\gamma $,n)$^{112m}$In by Gauss function

\begin{tabular}{|l|l|l|l|l|l|}
\hline Model& \textit{$\chi $}$^{2}$ & $S$& $E_{max}$, MeV&
$\Gamma $, MeV&
\textit{$\sigma $}$_{max}$,  mb \\
\hline ldm-1& 3.44& 883.0&15.7&3.9&180.0 \\
\hline ldm-2& 3.47&907.1& 15.7& 4.0&181.5 \\
\hline ldm-3& 3.31&895.7& 15.7& 4.0&181.7 \\
\hline ldm-4& 2.45&904.8& 15.8& 4.0&178.5 \\
\hline ldm-5& 3.67&914.0& 15.7& 3.9&188.3 \\
\hline \cite{12}& 1.46& 852.0& 15.7& 3.9&174.5 \\
\hline PL& 0.71&741.0& 15.9& 3.6&161.8 \\
\hline T& 1.69& 744.5&15.9& 3.6&166.8 \\
\hline
\end{tabular}

Table 4. The results of processing the peaks of cross section for
reaction $^{113}$In($\gamma $,n)$^{112m}$In by Breit-Wigner
function

\begin{tabular}{|l|l|l|l|l|l|}
\hline Model& \textit{$\chi $}$^{2}$ & $S$& $E_{max}$, MeV&
$\Gamma $, MeV&
\textit{$\sigma $}$_{max}$, mb \\
\hline ldm-1& 2.45&1575& 15.7&4.7&213.9 \\
\hline ldm-2&2.62& 1627& 15.7& 4.8&216.2 \\
\hline ldm-3&2.94& 1600& 15.7& 4.7&216.1 \\
\hline ldm-4&1.28& 1632& 15.8& 4.9&213.2 \\
\hline ldm-5&3.60& 1624& 15.7& 4.6&223.4 \\
\hline \cite{12}& 0.72& 1749& 15.7& 5.1&219.5 \\
\hline PL&0.47& 1243& 15.8& 4.2&190.1 \\
\hline T&0.85& 1277& 15.9& 4.1&198.6 \\
\hline
\end{tabular}

Table 5. The parameters of Breit-Wigner formula from fitting the
peaks of cross section for reaction $^{113}$In($\gamma
$,n)$^{112m}$In

\begin{tabular}{|l|l|l|l|}
\hline & \cite{12}& PL&T \\
\hline $y_{0}$& -20.65$\pm $7.2&15.81$\pm $9.0&11.44$\pm $7.3 \\
\hline $A$&1749$\pm $157&1243$\pm $168&1277$\pm $126 \\
\hline $x_{C}$& 15.7$\pm $0.1&15.8$\pm $0.1&15.9$\pm $0.1 \\
\hline $w$& 5.1$\pm $0.4&4.2$\pm $0.5&4.1$\pm $0.3 \\
\hline
\end{tabular}

\textbf{3. Cross section for reaction} $^{115}$\textbf{In($\gamma
$,n)}$^{114m}$\textbf{In}

On the yields \cite{14} in work \cite{15} was calculated cross
section of the reaction $^{115}$In($\gamma $,n)$^{114m}$In. Using
the package TALYS-1.6 we calculated the cross section for reaction
$^{115}$In($\gamma $,n)$^{114m}$In in the energy range of
10-25~MeV. The results obtained for five models of the density
levels of the nuclide (ldmodel 1-5) are shown in Fig. 3. For
ldmodel 1-3 values of cross section in the max are close
(42-45~mb), but differ from the experimental ones to 1.17 times.

The result of processing of peaks of the cross section energy
dependence for the reaction $^{115}$In($\gamma $,n)$^{114m}$In is
given in Table~6. Fitting the peak by Gaussian gives better
results than the fitting by Breit-Wigner formula.

\pdfximage width 177mm {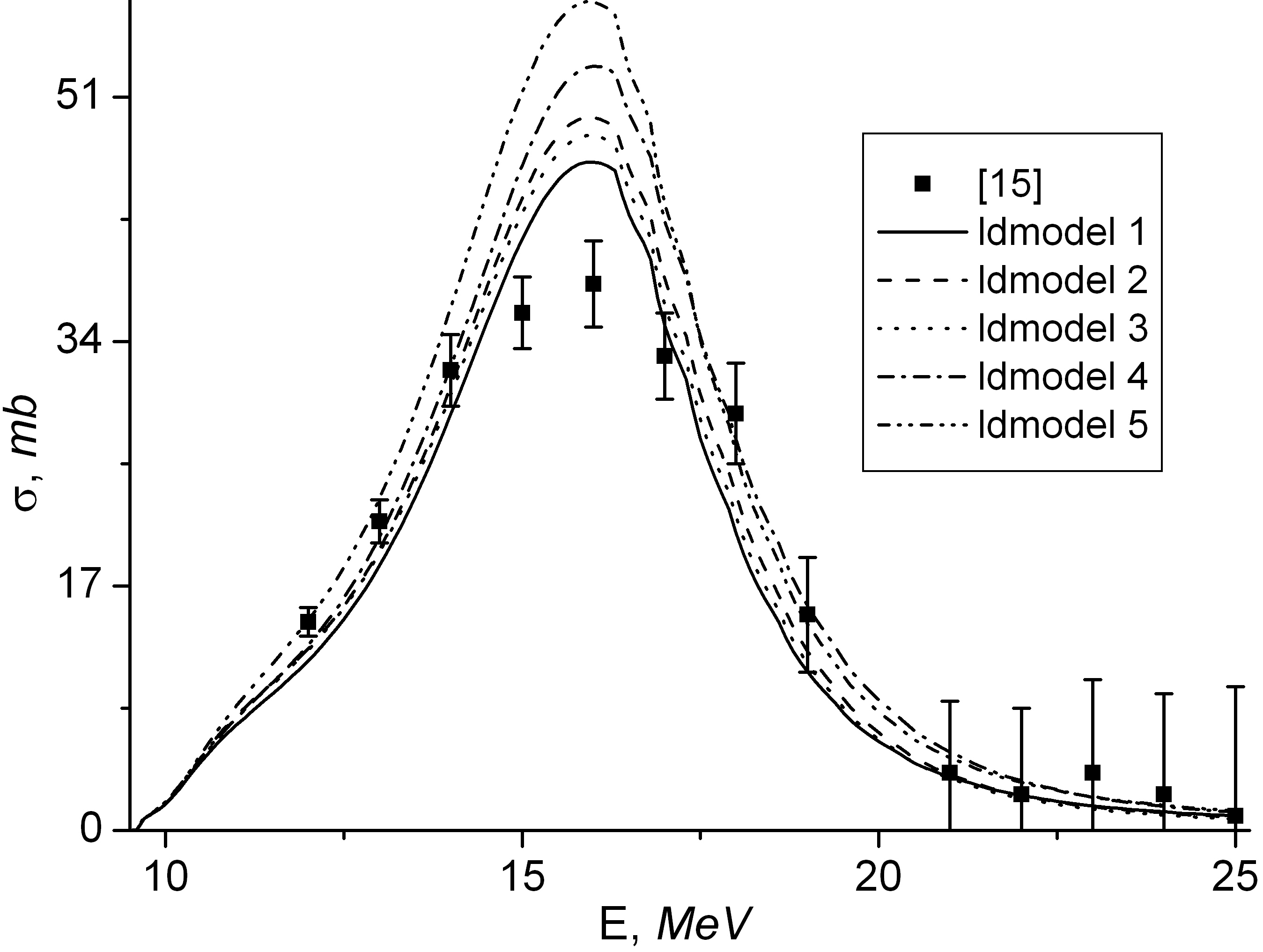}\pdfrefximage\pdflastximage

Fig.~3. Energy dependence of cross section of reaction
$^{115}$In($\gamma $,n)$^{114m}$In

Table 6. The results of processing the peaks of cross section for
reaction $^{115}$In($\gamma $,n)$^{114m}$In

\begin{tabular}{|l|l|l|l|l|l|}
\hline Model& \textit{$\chi $}$^{2}$ & $S$& $E_{max}$, MeV&
$\Gamma $, MeV& \textit{$\sigma $}$_{max}$, mb \\
\hline ldm-1&2.60& 203.9& 15.6& 3.8&42.3 \\
\hline ldm-2&2.74& 223.0& 15.7& 3.9&45.4 \\
\hline ldm-3&2.88& 216.2& 15.7& 3.9&44.2 \\
\hline ldm-4&2.64& 240.9& 15.8& 4.0&48.1 \\
\hline ldm-5& 3.46&255.4& 15.7& 3.9&52.5 \\
\hline \cite{15}& 2.37& 232.7& 15.7& 5.1&36.4 \\
\hline
\end{tabular}

\textbf{Conclusions}

The paper presents the results of experimental studies of the
isomeric relations of the yields for the reaction
$^{113}$In($\gamma $,n)$^{112m,g}$In, which recreated the energy
dependence of the cross section for the reaction
$^{113}$In($\gamma $,n)$^{112m}$In. Energy behavior of the cross
section for reaction has a characteristic shape of the giant
dipole resonance in the area of 15.8 MeV.

Experimental data cross section ($\gamma $,n)- reaction on indium
isotopes with the formation of isomers $^{112m}$In and $^{114m}$In
are compared with theoretical calculations in the package
TALYS-1.6. The analysis shows that both experimental and
theoretical calculations of the energy dependence of the cross
sections are described better by the Breit-Wigner function.

The obtained experimental results of the cross section for the
reaction $^{113}$In($\gamma $,n)$^{112m}$In can fill the nuclear
data bases of isomeric states which are used as constants in
nuclear applications, for example, for $\gamma $- activation
analysis.

This work was performed within the framework of the grant of the
Ministry of education and science of Ukraine on the topic of the
research work of the state registration number 0115U001098.

\end{document}